\documentclass[aps,prl,amsmath,amssymb,amsfonts,twocolumn,superscriptaddress,groupedaddress]{revtex4}
\usepackage{amsthm}
\usepackage{epsfig}
\usepackage{graphicx}
\usepackage{subfigure}
\usepackage{dcolumn}
\usepackage{bm}
\usepackage{braket}

\newcommand{\del}{\partial}

\newcommand{\bbR}{\mathbb{R}}

\newcommand{\veca}{\mathbf{a}}
\newcommand{\vecr}{\mathbf{r}}
\newcommand{\vecm}{\mathbf{m}}
\newcommand{\vecmbar}{\mathbf{\bar m}}
\newcommand{\vecp}{\mathbf{p}}
\newcommand{\vecsigma}{\boldsymbol{\sigma}}
\newcommand{\vecgamma}{\boldsymbol{\gamma}}

\begin{document}

\title{Spinor-helicity variables for cosmological horizons in de Sitter space}

\author{Adrian David}
\email{adrian.david@oist.jp}
\affiliation{Okinawa Institute of Science and Technology, 1919-1 Tancha, Onna-son, Okinawa 904-0495, Japan}
\author{Nico Fischer}
\email{nico.fischer@uni-jena.de}
\affiliation{Friedrich Schiller University Jena, 07737 Jena, Germany}
\author{Yasha Neiman}
\email{yashula@icloud.com}
\affiliation{Okinawa Institute of Science and Technology, 1919-1 Tancha, Onna-son, Okinawa 904-0495, Japan}

\date{\today}

\begin{abstract}
 We consider massless fields of arbitrary spin in de Sitter space. We introduce a spinor-helicity formalism, which encodes the field data on a cosmological horizon. These variables reduce the free S-matrix in an observer's causal patch, i.e. the evolution of free fields from one horizon to another, to a simple Fourier transform. We show how this result arises via twistor theory, by decomposing the horizon$\leftrightarrow$horizon problem into a pair of (more symmetric) horizon$\leftrightarrow$twistor problems.
\end{abstract}

\maketitle

\section{Introduction}

In field theory on flat spacetime, the S-matrix between past and future infinity is an object of fundamental importance. For \emph{massless} theories such as Yang-Mills and General Relativity (GR), the spinor-helicity formalism \cite{Kleiss:1985yh} has emerged as the ideal language \cite{Dixon:2013uaa} for studying the S-matrix (with the exception of some highly symmetric cases, in which twistor language is superior \cite{Mason:2009sa,ArkaniHamed:2009si,ArkaniHamed:2012nw}). Since our Universe appears to have a positive cosmological constant, it is of great theoretical interest to study the ``S-matrix'' in a static (i.e. observable) patch of de Sitter space, with an observer's past and future horizons in the roles of past/future infinity. So far, there's been remarkably little work on this problem. Instead, the main focus of theoretical attention in de Sitter space has been with correlations on its conformal boundary \cite{Maldacena:2002vr,Maldacena:2011nz,Arkani-Hamed:2017fdk}, which are unobservable in a true asymptotic de Sitter space (but become observable in approximate, temporary de Sitter scenarios such as inflation). 

In this letter, we take some first steps towards the de Sitter S-matrix. First, we encode the lightlike field data on a cosmological horizon in terms of spinor-helicity variables, equivalent to those introduced in \cite{Maldacena:2011nz} for the Poincare patch (see also the constructions for anti-de Sitter, in the Poincare patch \cite{Mirian} and in stereographic coordinates \cite{Nagaraj:2018nxq}). Then, in our main result, we relate the spinor-helicity variables associated with two cosmological horizons (and thus two Poincare patches), to obtain the free S-matrix in the static patch for massless fields of any spin. Our formalism and result provide a plausible starting point for efficiently including the effects of interactions in future work.

\section{Geometric setup}

De Sitter space is best described as a hyperboloid of unit spacelike radius embedded in flat 4+1d spacetime:
\begin{align}
 dS_4 = \left\{x^\mu \in \bbR^{1,4} \,|\ x_\mu x^\mu = 1 \right\} \ . \label{eq:dS}
\end{align}
We will use lightcone coordinates $x^\mu = (u,v,\vecr)$ for $\bbR^{1,4}$, where $\vecr$ is an $\bbR^3$ vector, and the metric is $dx_\mu dx^\mu = -dudv + \mathbf{dr}^2$. These coordinates are adapted to a de Sitter observer, whose initial and final horizons are defined by $(u=0,v<0)$ and $(u>0,v=0)$ respectively. The horizons' spatial section is the 2-sphere $S_2$ of unit vectors $\vecr^2 = 1$. The tangent space of this $S_2$ at a point $\vecr$ can be spanned by a complex null basis $(\vecm,\vecmbar)$:
\begin{align}
 \vecm\cdot\vecr = 0 \ ; \quad \vecm^2 = 0 \ ; \quad \vecm\times\vecmbar = -i\vecr \ . \label{eq:m}
\end{align}
This basis is defined up to phase rotations $(\vecm,\vecmbar)\rightarrow (e^{i\theta}\vecm,e^{-i\theta}\vecmbar)$, which describe $SO(2)$ rotations of the $S_2$ tangent space. In our setup, these rotations will play the role of the massless fields' little group.

Vectors in $dS_4$ are simply $\bbR^{1,4}$ vectors constrained to the tangent space of the hyperboloid \eqref{eq:dS}. Spinors in $dS_4$ can be constructed similarly from embedding-space spinors (see e.g. \cite{Neiman:2013hca}), but we will not need that construction here. For the statement of our main result, it will suffice to introduce the 2-component spinors $\psi^\alpha$ of spatial $SO(3)$ rotations. The antisymmetric metric on $SO(3)$ spinors is $\epsilon_{\alpha\beta}$, with inverse $\epsilon^{\alpha\gamma}\epsilon_{\beta\gamma} = \delta^\alpha_\beta$. We raise and lower indices via $\psi_\alpha = \epsilon_{\alpha\beta}\psi^\beta$. We denote the Pauli matrices by $\vecsigma^\alpha{}_\beta$. Spinors have a complex conjugation $\psi^\alpha\rightarrow\bar\psi^\alpha\rightarrow-\psi^\alpha$, under which $\epsilon_{\alpha\beta}$ is real but $\vecsigma^\alpha{}_\beta$ is imaginary. 

For the \emph{derivation} of our main result, we will also need the 4-component spinors of the $\bbR^{1,4}$ embedding space, i.e. the \emph{twistors} of $dS_4$. These can be constructed as pairs of $SO(3)$ spinors (see e.g. \cite{Neiman:2018ufb}):
\begin{align}
  Y^a = \begin{pmatrix} \lambda_\alpha \\ i\bar\mu^\alpha \end{pmatrix} \ , \label{eq:Y}
\end{align}
where the $i$ and complex conjugation on the second spinor are for later convenience. The $SO(1,4)$ spinor index $a$ is lowered via $Y_a = (-i\bar\mu^\alpha,\lambda_\alpha)$. Complex conjugation is inherited directly from that of the $SO(3)$ spinors. The $\bbR^{1,4}$ gamma matrices $\gamma_\mu = (\gamma_u,\gamma_v,\vecgamma)$ can be written in $2\times 2$ block notation as:
\begin{align}
 \begin{split}
   (\gamma_\mu)^a{}_b = \left(\begin{pmatrix} 0 & 0 \\ -\epsilon^{\alpha\beta} & 0 \end{pmatrix} , \begin{pmatrix} 0 & \epsilon_{\alpha\beta} \\ 0 & 0 \end{pmatrix} , \begin{pmatrix} -i\vecsigma_\alpha{}^\beta & 0 \\ 0 & -i\vecsigma^\alpha{}_\beta \end{pmatrix} \right) \ .
 \end{split} \nonumber
\end{align} 
We will sometimes omit both $SO(3)$ and $SO(1,4)$ spinor indices. In a product, this will imply bottom-to-top index contraction.

\section{Field data on the horizon}

We consider the free massless field equation for a totally-symmetric, double-traceless spin-$s$ gauge potential $h_{\mu_1\dots\mu_s}$ in $dS_4$ \cite{Fronsdal:1978vb}:
\begin{align}
 \begin{split}
   &\big(\Box + 2(s^2-1)\big) \phi_{\mu_1\dots\mu_s} - s\nabla_\rho\nabla_{(\mu_1}\phi^\rho_{\mu_2\dots\mu_s)} \\
   &\qquad + \frac{s(s-1)}{2}\nabla_{(\mu_1}\nabla_{\mu_2}\phi_{\mu_3\dots\mu_s)\nu}^\nu = 0 \ ,
 \end{split}
\end{align}
with a gauge symmetry $\delta\phi_{\mu_1\dots\mu_s} = \nabla_{(\mu_1}\Lambda_{\mu_2\dots\mu_s)}$ for totally-symmetric, traceless $\Lambda_{\mu_1\dots\mu_{s-1}}$.
The cases $s=0,1,2$ describe the confomally-coupled massless scalar, the Maxwell equations and linearized GR, respectively. In the scalar case, the field's value $\phi(u,0,\vecr)$ on e.g. the final horizon constitutes good boundary data for the field equation $(\Box - 2)\phi = 0$. For nonzero spin, good boundary data consists of one complex scalar component for the right-handed helicity, and its complex conjugate for the left-handed one; see e.g. \cite{Pritchard:1978ts,Bengtsson:1983pd,Ponomarev:2016lrm} for the standard construction in flat spacetime, and \cite{Penrose:1980yx} for a general discussion in terms of field strengths. In our present context, we can fix a gauge such that $\phi_{\mu_1\dots\mu_s}$ on the horizon has only spatial components $\phi_{i_1\dots i_s}$. Here, the $i_k$'s are $\bbR^3$ indices, which must be tangent to the $dS_4$ hyperboloid, and thus to the $S_2$ horizon section. The horizon boundary data is then given by the \emph{traceless part} of this $\phi_{i_1\dots i_s}$. Using the complex basis \eqref{eq:m} for the $S_2$ tangent space, we can reduce this traceless part to a pair of scalars:
\begin{align}
 \begin{split}
   \phi^{(s)}(u,\vecr;\vecm) &= m^{i_1}\dots m^{i_s}\phi_{i_1\dots i_s}(u,0,\vecr) \ ; \\
   \phi^{(-s)}(u,\vecr;\vecm) &= \bar m^{i_1}\dots\bar m^{i_s}\phi_{i_1\dots i_s}(u,0,\vecr) \ .
 \end{split} \label{eq:data_m}
\end{align}
These respectively describe fields of helicity $\pm s$, and carry weight $\pm s$ under the phase rotation $(\vecm,\vecmbar)\rightarrow (e^{i\theta}\vecm,e^{-i\theta}\vecmbar)$. The symplectic form for the horizon data \eqref{eq:data_m} reads:
\begin{align}
 \Omega[\delta\phi_1,\delta\phi_2] = \sum_{h = \pm s} \int \!du \int_{S_2} \!d^2\vecr\, \delta\phi_2^{(h)} \overleftrightarrow{\frac{\del}{\del u}} \delta\phi_1^{(-h)} \ , \label{eq:Omega}
\end{align}
where we sum over the two helicities $\phi^{(\pm s)}$ in the spinning case, or over just one helicity $\phi^{(0)}\equiv\phi$ in the scalar case. 

The boundary data on the \emph{initial} horizon can be encoded in the same way. Replacing the null time $u$ with $v$, and noticing that the helicity associated with $(\vecm,\vecmbar)$ is now reversed, we write:
\begin{align}
 \begin{split}
  \tilde\phi^{(s)}(v,\vecr;\vecm) &= \bar m^{i_1}\dots\bar m^{i_s}\phi_{i_1\dots i_s}(0,v,\vecr) \ ; \\
  \tilde\phi^{(-s)}(v,\vecr;\vecm) &= m^{i_1}\dots m^{i_s}\phi_{i_1\dots i_s}(0,v,\vecr) \ .
 \end{split} \label{eq:data_m_initial}
\end{align}
Finally, it's useful to define the gauge-invariant field strength data corresponding to the gauge potential data \eqref{eq:data_m},\eqref{eq:data_m_initial}:
\begin{align}
 C^{(\pm s)}(u,\vecr;\vecm) &= \frac{\del^s}{\del u^s}\,\phi^{(\pm s)}(u,\vecr;\vecm) \ ; \label{eq:data_C} \\
 \tilde C^{(\pm s)}(v,\vecr;\vecm) &= \frac{\del^s}{\del v^s}\,\tilde\phi^{(\pm s)}(v,\vecr;\vecm) \ . \label{eq:data_C_initial}
\end{align}

\section{The S-matrix problem}

For our purposes, the S-matrix problem in de Sitter space is to relate the gauge-invariant field data \eqref{eq:data_C} on the final horizon to the corresponding data \eqref{eq:data_C_initial} on the initial one. This statement of the problem, which will be more convenient for us, is slightly more general than what is usually termed the S-matrix. Usually, one would relate the \emph{quantum states} obtained by acting with the fields on some vacuum; by focusing on the fields themselves, we avoid committing to a particular vacuum state. We will ignore here any subtleties related to zero-frequency modes, i.e. to the horizons' lower-dimensional boundaries (either at asymptotic infinity or at the horizons' $S_2$ intersection). In other words, we will be dealing with the ``hard part'' of the S-matrix.

For free fields, one can find the S-matrix by ``brute force'', using the general technique for linear hyperbolic equations. Essentially, the value of a massless field at some final horizon point is determined by the intersection of that point's past lightcone with the initial horizon. Thus, for e.g. the scalar field, we have:
\begin{align}
 \phi(u,0,\vecr) = \frac{1}{\pi u}\int_{S_2}\!d^2\vecr' \left.\frac{\del\phi(0,v,\vecr')}{\del v} \right|_{v = 2(\vecr\cdot\vecr' - 1)/u} \ ,  \label{eq:brute_force}
\end{align}
which can be obtained from the general formula:
\begin{align}
 \phi(u,0,\vecr) = \int\!dv\int_{S_2}\!d^2\vecr'\,\phi(0,v,\mathbf{r'}) \overleftrightarrow{\frac{\del}{\del v}} G(u,0,\vecr;0,v,\mathbf{r'}) \ , \nonumber
\end{align}
in which $G(x^\mu;x'^\mu) = (-1/4\pi)\delta(x_\mu x'^\mu - 1)\theta(u-u')$ is a causal Green's function in $dS_4$. For the analogous general treatment of nonzero spin, see \cite{Penrose:1980yx}. In fact, the end result \eqref{eq:brute_force} holds not only in the static patch $(u>0,v<0)$, but also for the horizons' entire extent $u,v\in\bbR$, which includes the antipodal patch $(u<0,v>0)$. In the rest of the letter, we'll avoid specifying the range of $u,v$, and our formulas will apply equally well to both $(u>0,v<0)$ and $u,v\in\bbR$. While the $(u>0,v<0)$ case is linked more directly to observable physics, our formulas ``live more naturally'' in the more global context $u,v\in\bbR$.

Despite its simplicity, eq. \eqref{eq:brute_force} is not quite satisfactory. Since it doesn't make explicit contact with the horizons' symmetries, it is unlikely as a useful starting point for interacting calculations. 

What, then, are the relevant symmetries? Naively, they are the subgroup of $dS_4$ isometries that preserves both horizons. These are the static-patch time translations $(u,v)\rightarrow (e^t u,e^{-t}v)$, and the $SO(3)$ rotations of $\vecr$. These symmetries encourage one to work in terms of frequencies and spherical harmonics. The S-matrix for the free scalar in this basis was found in \cite{Hackl:2014txa,Halpern:2015zia}. However, spherical harmonics are rather unpleasant, so the generalization to interacting theories again does not seem promising. Below, we will describe a different basis for the S-matrix, which replaces spherical harmonics with plane waves, by fixing only one horizon at a time. 

\section{Poincare momentum and spinor-helicity}

Instead of fixing \emph{both} horizons, let us consider the residual symmetry from fixing just e.g. the final one. This is the symmetry of the Poincare patch: the translations, rotations and dilatations of $\bbR^3$. On the horizon, the rotations act on $\vecr\in S_2$ in the obvious way, the dilatations rescale $u$, while a translation by a vector $\veca$ shifts the lightrays according to $u\rightarrow u - 2\veca\cdot\vecr$. A fixed \emph{momentum} $\vecp$ with respect to these translations describes two modes on the horizon:
\begin{align}
 &\text{Positive frequency:} & &\delta^2\!\left(\vecr,+\frac{\vecp}{|\vecp|}\right) e^{-i|\vecp|u/2} \ ; \\
 &\text{Negative frequency:} & &\delta^2\!\left(\vecr,-\frac{\vecp}{|\vecp|}\right) e^{+i|\vecp|u/2}\ . \label{eq:negative_freq}
\end{align}
These modes are waves with frequency $\pm|\vecp|/2$ with respect to the null time $u$, supported on an antipodal pair of lightrays $\vecr = \pm\vecp/|\vecp|$.

Let us now define spinor-helicity variables $\lambda_\alpha,\bar\lambda_\alpha$ as the spinor square root of $\vecp$, such that $\vecp = \bar\lambda\vecsigma\lambda$. The corresponding positive-frequency mode will be a wave supported at $\vecr = (\bar\lambda\vecsigma\lambda)/(\bar\lambda\lambda)$, with frequency $\bar\lambda\lambda/2$ with respect to $u$. Moreover, we can use $\vecm = i(\lambda\vecsigma\lambda)/(\sqrt{2}\bar\lambda\lambda)$ and its complex conjugate $\vecmbar$ as the complex null basis \eqref{eq:m} for the $S_2$ tangent space. Thus, we package the positive-frequency part of the horizon field data \eqref{eq:data_m} into spinor functions $f^{(\pm s)}(\lambda_\alpha,\bar\lambda_\alpha)$ as follows:
\begin{align}
 f^{(\pm s)}(\lambda_\alpha,\bar\lambda_\alpha) = \int\!du\,e^{i(\bar\lambda\lambda)u/2}\,
   \phi^{(\pm s)}\!\left(u,\frac{\bar\lambda\vecsigma\lambda}{\bar\lambda\lambda} ; \frac{i\lambda\vecsigma\lambda}{\sqrt{2}\bar\lambda\lambda} \right) \ . \label{eq:spin_hel}
\end{align}
These spinor functions have the following manifest symmetries:
\begin{itemize}
 \item By construction, they have momentum $\bar\lambda\vecsigma\lambda$ under the Poincare-patch translations $u\rightarrow u - 2\mathbf{a}\cdot\vecr$.
 \item Under Poincare-patch dilatations, i.e. static-patch time translations $(u,v)\rightarrow (e^t u,e^{-t}v)$, they scale as $f^{(\pm s)}(\lambda,\bar\lambda) \rightarrow e^t f^{(\pm s)}(e^{t/2}\lambda,e^{t/2}\bar\lambda)$. 
 \item The field's helicity $\pm s$ is encoded in the scaling relation $f^{(\pm s)}(e^{i\theta/2}\lambda,e^{-i\theta/2}\bar\lambda) = e^{\pm is\theta}f^{(\pm s)}(\lambda,\bar\lambda)$.
\end{itemize}
In fact, along with the more obvious $SO(3)$ rotations, these symmetries \emph{uniquely determine} the encoding \eqref{eq:spin_hel} of horizon data into spinor functions, up to a prefactor that can only depend on helicity. As a corollary, we conclude that this spinor-helicity formalism must coincide with the one constructed from a different point of view in \cite{Maldacena:2011nz}.

As is frequently useful in the spinor-helicity formalism \cite{Britto:2005fq}, we can analytically continue to complex momenta by making the two spinors $(\lambda,\bar\lambda)$ independent. As a special case, by analytically continuing $(\lambda,\bar\lambda)\rightarrow(i\lambda,i\bar\lambda)$, we obtain the \emph{negative-frequency} modes:
\begin{align}
 f^{(\pm s)}(i\lambda,i\bar\lambda) = \left(f^{(\mp s)}(\lambda,\bar\lambda) \right)^* \ . \label{eq:continue}
\end{align}
In these variables, the field's symplectic form \eqref{eq:Omega} reads simply:
\begin{align}
 \begin{split}
   &\Omega[\delta f_1,\delta f_2] = \frac{1}{i}\sum_{h = \pm s} \int \!\frac{d^2\lambda\,d^2\bar\lambda}{(2\pi i)^2} \\
   &\qquad \left( \delta f^{(h)}_1(\lambda,\bar\lambda)\,\delta f^{(-h)}_2(i\lambda,i\bar\lambda) \,-\, (1\leftrightarrow 2) \right) \ ,
 \end{split}
\end{align}
where $d^2\lambda$ is the spinor measure $\epsilon_{\alpha\beta}d\lambda^\alpha d\lambda^\beta/2$.

The field data \eqref{eq:data_m_initial} on the \emph{initial} horizon can be treated in the same way. Thus, the positive-frequency modes are encoded as:
\begin{align}
 \tilde f^{(\pm s)}(\mu^\alpha,\bar\mu^\alpha) = \int\!dv\,e^{i(\bar\mu\mu)v/2}\,
   \tilde\phi^{(\pm s)}\!\left(v,\frac{\bar\mu\vecsigma\mu}{\bar\mu\mu} ; \frac{i\mu\vecsigma\mu}{\sqrt{2}\bar\mu\mu} \right) \ . \label{eq:spin_hel_initial}
\end{align}
The negative-frequency modes can again be obtained by analytic continuation, as in \eqref{eq:continue}. Note that $\mu^\alpha$ is now the square root of momentum in a \emph{different} Poincare coordinate patch -- the one associated with the initial horizon.

\section{The free S-matrix}

In the framework of the previous section, finding the S-matrix means relating the spinor functions $f^{(\pm s)}(\lambda,\bar\lambda)$ and $\tilde f^{(\pm s)}(\mu,\bar\mu)$. Our letter's main result (to be derived in the next section) is that the free-field answer is simply a Fourier transform:
\begin{align}
 f^{(\pm s)}(\lambda_\alpha,\bar\lambda_\alpha) = \int \frac{d^2\mu\,d^2\bar\mu}{(2\pi i)^2}\,\tilde f^{(\pm s)}(\mu^\alpha,\bar\mu^\alpha)\,
   e^{i(\lambda_\alpha\mu^\alpha + \bar\lambda_\alpha\bar\mu^\alpha)} \ . \label{eq:result}
\end{align}
This formula can be unpacked explicitly in terms of the horizon field data \eqref{eq:data_C}-\eqref{eq:data_C_initial}, giving:
\begin{align}
  &C^{(\pm s)}(u,\vecr;\vecm) = \frac{2^s}{\pi u^{2s+1}} \label{eq:result_explicit} \\
  &\ \times \int_{S_2}\!d^2\vecr'\left(1 - \vecr\cdot\vecr' \right)^s \left.\frac{\del\tilde C^{(\pm s)}(v,\vecr';\vecm')}{\del v} \right|_{v = 2(\vecr\cdot\vecr' - 1)/u} \ , \nonumber
\end{align}
which reproduces the spin-0 result \eqref{eq:brute_force} as a special case. The phase of the null tangent vector $\vecm'$ in \eqref{eq:result_explicit} is fixed via:
\begin{align}
 2\vecm\cdot\vecm' = 1 - \vecr\cdot\vecr' \ . \label{eq:m'_phase}
\end{align} 

The explicit S-matrix \eqref{eq:result_explicit} can be derived from the main result \eqref{eq:result} and the definitions \eqref{eq:data_C}-\eqref{eq:data_C_initial},\eqref{eq:spin_hel},\eqref{eq:spin_hel_initial}, via mostly straightforward integrals. The two non-trivial ``tricks'' are:
\begin{enumerate}
	\item When integrating over $\bar\lambda\lambda$ to invert the transform \eqref{eq:spin_hel}, it helps to also average over the phase of $\lambda_\alpha$, using the known weight of $f^{(\pm s)}$ under such phase rotations. We then have an integral over both magnitudes and phases, which becomes an integral $dz d\bar z$ over the complex plane.
	\item Conversely, when integrating over $\mu^\alpha,\bar\mu^\alpha$ in \eqref{eq:result}, it helps to localize the integral on values of $\mu^\alpha$ with an overall phase given by \eqref{eq:m'_phase}, where $\vecr' \equiv (\bar\mu\vecsigma\mu)/(\bar\mu\mu)$ and $\vecm' \equiv (i\mu\vecsigma\mu)/(\sqrt{2}\bar\mu\mu)$.
\end{enumerate}

\section{Twistorial derivation}

To prove our S-matrix formula \eqref{eq:result}, we will relate it to a picture that is covariant under the full $SO(1,4)$ de Sitter group. First, we'll rewrite our transforms \eqref{eq:spin_hel},\eqref{eq:spin_hel_initial} between spinor-helicity functions and horizon data in $SO(1,4)$-covariant language. For this, we combine $\lambda_\alpha$ and $i\bar\mu^\alpha$ into a \emph{twistor}, i.e. an $SO(1,4)$ spinor $Y^a$, as in \eqref{eq:Y}. The final horizon's spinor functions \eqref{eq:spin_hel} can now be treated as functions of the twistor variables $(Y,\bar Y)$, which just happen to depend only on the $(\gamma_u Y,\gamma_u\bar Y)$ components, i.e. the ones containing $(\lambda,\bar\lambda)$:
\begin{align}
 f(Y,\bar Y) &= \int\!du\,e^{i(\bar Y\!\gamma_u\!Y)u/2}\,
   \phi\!\left(u,\frac{\bar Y\!\gamma_u\vecgamma Y}{i\bar Y\gamma_u Y} ; \frac{Y\!\gamma_u\vecgamma Y}{\sqrt{2}\,\bar Y\gamma_u Y} \right) \ , \nonumber
\end{align}
where we removed the helicity superscripts to save space. The \emph{initial} horizon's spinor functions \eqref{eq:spin_hel_initial} are now given by the exact same formulas, but with $u$ replaced everywhere by $v$. Of course, this replacement amounts to interchanging the null axes in $\bbR^{1,4}$ that define the two horizons. 

For the second step of our proof, recall that free massless fields in $dS_4$ (or in any conformally flat 4d spacetime) can be encoded, via the Penrose transform \cite{Penrose:1986ca,Ward:1990vs}, as \emph{holomorphic twistor functions} $F^{(\pm s)}(Y^a)$, with no dependence on the complex conjugate $\bar Y^a$. The de Sitter group $SO(1,4)$ (and, in fact, the entire conformal group $SO(2,4)$) is realized on these functions by linear transformations of the twistor argument $Y^a$. In particular, the Poincare-patch translations $u\rightarrow u - 2\veca\cdot\vecr$ and dilatations $(u,v)\rightarrow (e^t u,e^{-t}v)$ are realized respectively as:
\begin{align}
 F^{(\pm s)}(\lambda_\alpha,i\bar\mu^\alpha) \, &\rightarrow \, F^{(\pm s)}\big(\lambda_\alpha, \, i\bar\mu^\alpha\! + i(\veca\cdot\vecsigma)^{\alpha\beta}\lambda_\beta \big) \ ; \nonumber \\
 F^{(\pm s)}(\lambda_\alpha,i\bar\mu^\alpha) \, &\rightarrow \, F^{(\pm s)}(e^{t/2}\lambda_\alpha, \, ie^{-t/2}\bar\mu^\alpha) \ ,
\end{align}
along with the obvious action of $SO(3)$ rotations. Finally, the field's helicity is encoded in the twistor function's degree of homogeneity:
\begin{align}
 \lambda_\alpha\frac{\del F^{(\pm s)}}{\del\lambda_\alpha} + \bar\mu^\alpha\frac{\del F^{(\pm s)}}{\del\bar\mu^\alpha} = Y^a\frac{\del F^{(\pm s)}}{\del Y^a} = -2 \pm 2s \ .
\end{align}
We can now see that the \emph{Fourier transform} of the twistor function $F^{(\pm s)}(\lambda,i\bar\mu)$ with respect to $\bar\mu^\alpha$ has the defining symmetries of the spinor-helicity function $f^{(\pm s)}(\lambda,\bar\lambda)$ on the final horizon! Therefore, up to a prefactor that can be absorbed into the definition of $F^{(\pm s)}(\lambda,i\bar\mu)$, we identify:
\begin{align}
 f^{(\pm s)}(\lambda,\bar\lambda) = \int \frac{d^2\bar\mu}{2\pi}\,F^{(\pm s)}(\lambda,i\bar\mu)\,e^{i\bar\lambda\bar\mu} \ , \label{eq:final_half_Fourier}
\end{align}
or, in $SO(1,4)$-covariant notation:
\begin{align}
 \begin{split}
   f(Y,\bar Y) = \int \frac{dZ\gamma_v dZ}{4\pi}\,&F(\gamma_v\gamma_u Y + \gamma_u\gamma_v Z)
     \, e^{i\bar Y\gamma_u\gamma_v Z} \ ,
 \end{split} \nonumber
\end{align}
where the helicity superscripts are again omitted. But now, by $SO(1,4)$ covariance, upon interchanging $u\leftrightarrow v$, we must get the spinor-helicity functions $\tilde f^{(\pm s)}(Y,\bar Y)$ for the \emph{initial} horizon! Back in $SO(3)$ spinor notation, this last statement reads:
\begin{align}
 \tilde f^{(\pm s)}(\mu,\bar\mu) = -\int\frac{d^2\lambda}{2\pi}\,F^{(\pm s)}(\lambda,i\bar\mu)\,e^{-i\lambda\mu} \ . \label{eq:initial_half_Fourier}
\end{align}
Combining the two ``half''-Fourier transforms \eqref{eq:final_half_Fourier}-\eqref{eq:initial_half_Fourier}, we obtain the free S-matrix \eqref{eq:result}.

\section{Conclusion}

In this letter, we applied a spinor-helicity formalism to the horizon data of massless fields in de Sitter space, and derived an elegant expression for the free S-matrix of massless fields with any spin. We extended the known analogies \cite{Maldacena:2011nz} with the spinor-helicity formalism in flat spacetime, in particular establishing a ``half''-Fourier transform relation between spinor-helicity and twistor functions. We also went beyond \cite{Maldacena:2011nz} by working with two different horizons, each of which has a different notion of momentum. Despite this seeming complication, we saw that the momenta on the two horizons are related by a simple Fourier transform \eqref{eq:result}, once we consider their spinor square root. This Fourier-transform relationship can also be understood as a change of basis in the Hilbert space of a massless particle on the (complexified) 3d conformal boundary \cite{Neiman:2018ufb}. 

Having found the free S-matrix in the spinor-helicity basis, we obtained an explicit expression \eqref{eq:result_explicit} in terms of horizon field data for any spin. More importantly, there is now hope that perturbation theory for interacting fields in the de Sitter causal patch can be constructed with relative ease. In the future, we would like to explore this possibility, in particular for Yang-Mills and GR. In addition, we expect that the language developed here can be fruitfully applied to higher-spin gravity \cite{Vasiliev:1995dn,Vasiliev:1999ba}, the only known working model \cite{Anninos:2011ui} for dS/CFT holography in 3+1 dimensions.

\section*{Acknowledgements}		

This work was supported by the Quantum Gravity and Mathematical \& Theoretical Physics Units of the Okinawa Institute of Science and Technology Graduate University (OIST). NF's work was carried out during an internship at OIST.

\end{document}